\pdfoutput=1
\documentclass[type=preprint,linenumbering=off]{sphenix}
\usepackage{amsmath,amssymb}
\usepackage{lipsum}
\usepackage{cleveref}
\crefrangeformat{figure}{Figs. #3#1#4--#5#2#6}
\usepackage[labelfont=normalfont]{subcaption}
\usepackage{cprotect}

\usepackage{orcidlink}

\newcommand{\Npart}{\mbox{$N_{\mathrm{part}}$}\xspace}

\newcommand{\snn}{\mbox{$\sqrt{s_{_{NN}}}$}\xspace}

\def\GeV{\ifmmode {\mathrm{\ Ge\kern -0.1em V}}\else
                   \textrm{Ge\kern -0.1em V}\fi \xspace}%
\newcommand{\sqsntwo}{\mbox{$\sqrt{s_{_{NN}}}=200$~\GeV}\xspace}

\newcommand{\PbPb}{\mbox{Pb+Pb}\xspace}

\newcommand{\auau}{\mbox{Au+Au}\xspace}

\def\pt{\mbox{$p_T$}\xspace}
\def\detdeta{\mbox{$dE_{T}/d\eta$}\xspace}

\def\MeV{\ifmmode {\mathrm{\ Me\kern -0.1em V}}\else
                   \textrm{Me\kern -0.1em V}\fi \xspace}%

\begin{document}

\title{Measurement of the transverse energy density in \auau{} collisions at \sqsntwo with the sPHENIX detector}

\author{sPHENIX Collaboration\footnote{See the appendix for the list of collaboration members}}

\date{\today}

\doctag{sPH-BULK-2025-02}
\docdoi{10.1103/h8d5-swg6}

\maketitle

\begin{abstract}
This paper reports measurements of the transverse energy per unit pseudorapidity (\detdeta) produced in \auau collisions at \sqsntwo, performed with the sPHENIX detector at the Relativistic Heavy Ion Collider (RHIC). The results cover the pseudorapidity range $\left|\eta\right| < 1.1$ and constitute the first such measurement performed using a hadronic calorimeter at RHIC. Measurements of \detdeta are presented for a range of centrality intervals and the average \detdeta as a function of the number of participating nucleons, \Npart, is compared to a variety of Monte Carlo heavy-ion event generators. The results are in agreement with previous measurements at RHIC, and feature an improved granularity in $\eta$ and improved precision in low-\Npart events.

\end{abstract}

\section{Introduction}
\label{sec:introduction}
Heavy-ion collisions produce hot and dense matter consisting of deconfined quarks and gluons known as the quark-gluon plasma (QGP)~\cite{Busza_2018} and allow for the study of fundamental aspects of the strong nuclear force. 
Measurements of the bulk properties of the final state, after the full evolution of the QGP, can be used to constrain its characteristics~\cite{Romatschke_2017}. 
An important quantity to study such properties is the transverse energy ($E_{T}$) per unit pseudorapidity ($\eta$), \detdeta, which is sensitive to the energy carried along the longitudinal direction, providing essential information related to the initial state that is propagated through subsequent hydrodynamic evolution of the QGP.

Early measurements of \detdeta performed by the PHENIX~\cite{PHENIX:2001kdi,PHENIX:2004vdg} and STAR~\cite{STAR_detdeta} experiments at the Relativistic Heavy Ion Collider (RHIC) were used to infer that the Bjorken energy densities~\cite{bjorken_energy_density} reached in \auau{} collisions at nucleon-nucleon center-of-mass energy of \sqsntwo are greater than 5~\GeV/fm$^3$ at 1~fm/c after the collision. 
These measurements demonstrated that a necessary pre-condition for QGP formation had been met, as these energy densities were above the Lattice Quantum Chromodynamics (QCD) predictions for the transition to a QGP~\cite{Aoki:2006we}. 
Similar measurements of \detdeta, also referred to as transverse energy density, have been performed at the Large Hadron Collider (LHC) in \PbPb{} collisions at higher energies~\cite{alice_detdeta,cms_pbpb_detdeta}. 
The centrality dependence of \detdeta, and its interpretation in terms of geometric quantities, such as the relationship to the average number of nucleon participants (\Npart), have also been quantified by these experiments to test models of particle production in large collision systems~\cite{Wang_Gyulassy_2001,Kharzeev:2000ph,Eremin:2003qn,PHENIX_npq_scaling_detdeta}. 

sPHENIX is a new detector located at RHIC, designed to provide qualitatively new capabilities for physics measurements in heavy-ion collisions~\cite{sPHENIX_proposal,sphenix_predictions}. The sPHENIX calorimeter system is composed of the Electromagnetic Calorimeter (EMCal) and a two-layer Hadronic Calorimeter (HCal), each providing a large acceptance of $\left|\eta\right| < 1.1$ with full azimuthal coverage. The sPHENIX HCal is the first mid-pseudorapidity HCal at RHIC.
The sPHENIX calorimeter system is purpose-built for high-precision measurements of jets, hadrons, and photons in heavy-ion collisions, and is also well-suited for measurements of global event properties such as \detdeta. The \detdeta measurement is a key benchmark for the sPHENIX calorimeter performance and provides confidence in the sPHENIX EMCal and HCal energy reconstruction for future jet measurements.
Furthermore, a detailed study of the underlying event in \auau{} collisions, and how it manifests in the sPHENIX calorimeter system, is essential for the development of the sPHENIX jet physics program~\cite{Hanks:2012wv}.

This paper reports a measurement of \detdeta in \auau collisions at \snn = 200~\GeV{} with the sPHENIX detector using the data collected during Run 2024 at RHIC. 
The results are presented as functions of $\eta$ and in different \auau collision centrality intervals. The \detdeta measurements are scaled by the average number of participant-nucleon pairs ($0.5$\Npart) and presented as a function of \Npart to inform particle-production models. These are compared to a selection of Monte Carlo (MC) heavy-ion event generators, \verb|AMPT|~\cite{ampt}, \verb|EPOS4|~\cite{epos}, and \verb|HIJING|~\cite{hijing}, as well as previous measurements of \detdeta at this collision energy~\cite{PHENIX:2004vdg,STAR_detdeta}. 
\section{sPHENIX detector}
\label{sec:sphenix_info}
The sPHENIX detector is the only large, general-purpose detector built at a hadron collider in the last decade. sPHENIX began its commissioning period in preparation for physics data taking in June 2023, and has now collected high-quality physics data from RHIC running in 2024. sPHENIX has multiple detector subsystems for carrying out a broad physics program including: (i) multiple detectors at forward rapidity to characterize the collision event, (ii) a four-component precision tracking detector system, and (iii) full coverage electromagnetic and hadronic calorimetry.
A detailed description of the sPHENIX detector is available here~\cite{sphenix_tdr}. This analysis uses a subset of the  collision-characterization detectors and the full calorimeter system, which are described in more detail below.

The Minimum Bias Detector (MBD) provides the collision vertex position and centrality categorization. The MBD is located at forward rapidity, $3.51 < |\eta| < 4.61$, on both sides of the interaction point (IP), surrounding the beam pipe, and comprises 128 channels of photomultiplier tubes (PMTs). The MBD is used for triggering on hadronic interactions providing a selection of minimum-bias (MB) events in heavy-ion collisions and for the collision z-vertex position reconstruction via the difference in arrival time of signals in the PMTs on both sides. The vertex position resolution in central Au+Au events is better than 1~cm. The MBD PMT charge, or PMT signal amplitude, is calibrated in terms of minimum ionizing particle (MIP) counts and is used to characterize the multiplicity of the collision. The MBD was previously used in PHENIX~\cite{Ikematsu:1998fm,phenix_mbd_nim} at a different position, $3.0 < |\eta| < 3.9$, and is included in sPHENIX with new readout and trigger electronics. 

The Zero Degree Calorimeters (ZDCs)~\cite{ADLER2001488} are located on both sides of the IP, at a distance of 18~m from the IP. These are sampling hadronic calorimeters composed of tungsten alloy and 
optical fibers. The fibers transmit Cherenkov light generated from charged particles traversing the detector volume to PMTs.   
The ZDC is incorporated into the MB event selection criteria to separate inelastic hadronic interaction events from beam backgrounds and other processes. Each side of the ZDC is calibrated by requiring that the single neutron peak in the measured energy distribution, summed from that deposited in its three compartments, is at a nominal value of 100~\GeV.

The calorimeter system comprises three radially concentric, cylindrical layers: the EMCal and the HCal, which includes the Inner Hadronic Calorimeter (IHCal), inside the bore of the 1.4~Tesla superconducting magnet, and the Outer Hadronic Calorimeter (OHCal) outside the magnet. Within the calorimeter system, the EMCal (OHCal) is located closest (furthest) from the interaction point. These detectors have full azimuthal coverage and extend $|\eta| < 1.1$ in pseudorapidity.

The EMCal~\cite{EM_test_beam,emcal_fudan,EMCal_HCal_test_beam} is approximately 20 radiation lengths deep and is designed to measure photons, electrons, and positrons via their electromagnetic showers. The EMCal is also approximately 0.8 hadronic interaction lengths deep and therefore also sees a significant amount of hadronic shower energy on average. The  EMCal is a sampling calorimeter composed of blocks filled with tungsten powder absorber and scintillating fibers, each subtending $\Delta \eta \times \Delta \phi = 0.025 \times 0.025$. The light from the scintillating fibers is collected by light guides and measured with silicon photo-multipliers (SiPMs). The resolution of the EMCal to electromagnetic showers has been determined using beam tests which report a resolution to electrons of $2.8\% \oplus 15.5\%/\sqrt{E}$~\cite{EM_test_beam,EMCal_HCal_test_beam}.

The sPHENIX HCal system~\cite{EMCal_HCal_test_beam} is designed to measure hadronic showers, with the full EMCal + HCal system totaling nearly five hadronic interaction lengths. Both the IHCal and OHCal are sampling calorimeters comprising aluminum (Inner) or steel (Outer) absorbing plates and scintillating tiles arranged into towers read out with granularity $\Delta \eta \times \Delta \phi = 0.1 \times 0.1$. Within the HCal towers, which are projective in $\eta$, the scintillating tiles are set at an angle offset relative to the direction pointing to the center of the detector in order to reduce the number of traversing particles that do not interact with the active volumes of these calorimeters. The scintillation light from these tiles is converted and captured by embedded wavelength-shifting fibers and then detected by SiPMs. The full calorimeter system (EMCal + HCal) resolution for hadronic showers determined from the energy resolution for hadrons in beam tests using the combined EMCal plus HCal system is $13.5\% \oplus 64.9\%/\sqrt{E}$~\cite{EMCal_HCal_test_beam}.

Signals produced by the EMCal and HCal SiPMs and MBD and ZDC PMTs are digitized and read out using a common set of calorimeter electronics. These analog signals are digitized with 14-bit precision at a frequency 56.4~MHz, which is six times the RHIC bunch crossing rate. In the case of the EMCal and HCal subsystems, digitized waveforms peaking above a hardware-level zero suppression threshold, defined by 2$\sigma$ of the calorimeter pedestal noise at the start of Run 2024, are read out recording the full 12-sample waveform, otherwise only two summary pre- and post-rise samples are saved. For the MBD and ZDC readout, no hardware-level zero suppression is employed and all time samples of the digitized waveform are recorded.
\section{Transverse energy measurement}
\label{sec:analysis}
\subsection{Event reconstruction}

\begin{figure}
    \centering
        \includegraphics[height=3.5in]{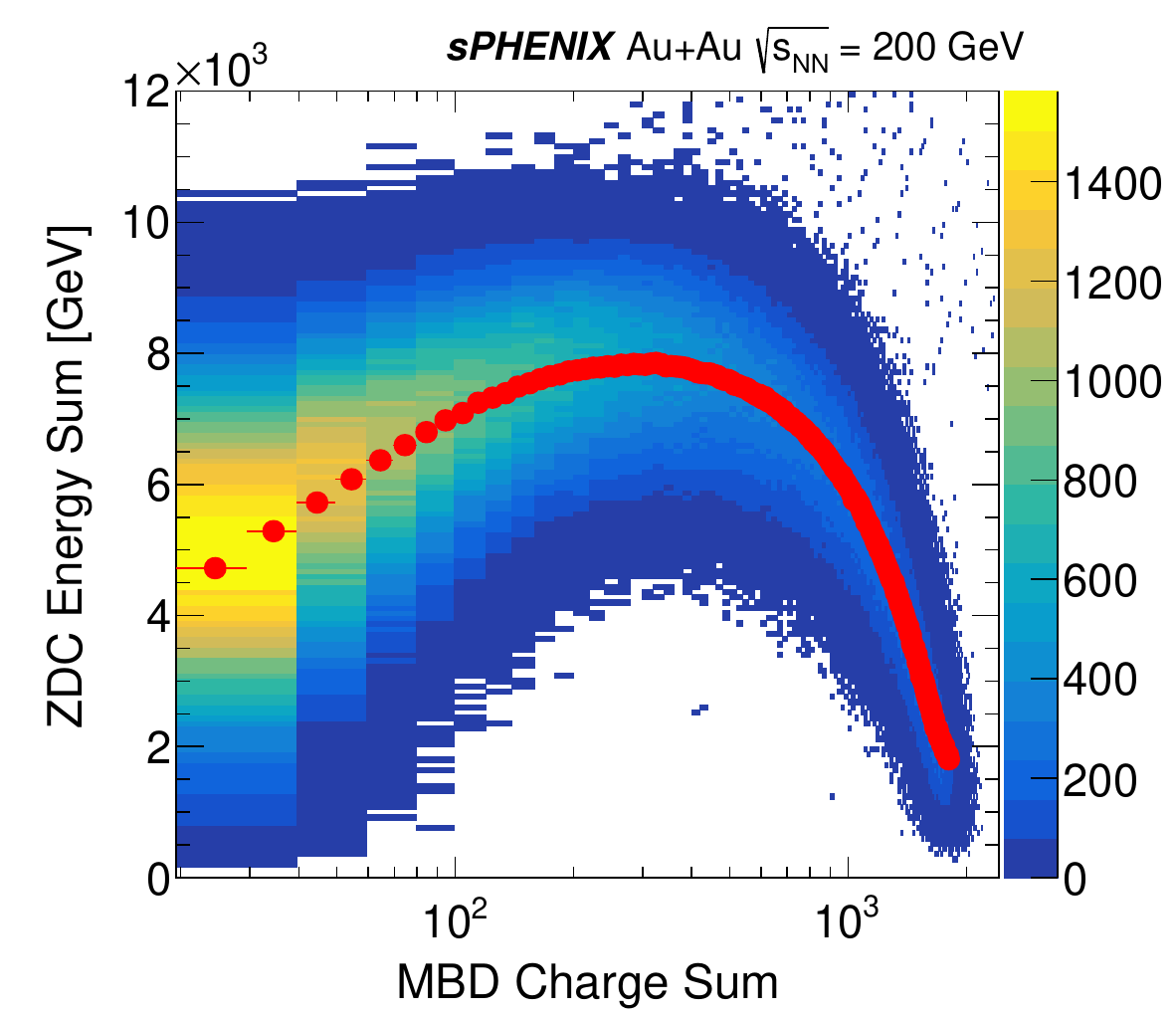}
        \caption{The correlation between the total energy in ZDCs and the MBD charge sum is shown.   The MBD charge sum is in units of calibrated MIPs.    The red points indicate the average ZDC energy as a function of MBD charge sum.}
        \label{fig:mbd_zdc}
\end{figure}

This analysis uses a small set of runs recorded during low-luminosity \auau running in RHIC Run 2024, originally intended to commission the sPHENIX tracking detectors. Pile-up effects, even in the most central events, are negligible for the runs used in this analysis. The MBD, ZDC, and calorimeters were in normal physics-taking mode and the magnetic field was at its nominal operating point of 1.4~T. 
Collision events were accepted using a hardware trigger which required at least two PMTs fired on each side of the MBD. In the offline analysis, a set of MB selection criteria has been applied using signals from both the MBD and ZDC detectors to remove events consistent with beam-related backgrounds and non-hadronic interactions. Figure~\ref{fig:mbd_zdc} shows the correlation of energy in the ZDCs to the MBD charge sum for events passing the MB selection criteria.
The MBD charge sum is the sum of charge from all PMTs on both sides of the MBD, and is calibrated such that one MIP is set at unity.
A characteristic decrease in the total ZDC energy is observed at both high and low MBD charge, corresponding to very central collisions (where only a small number of spectator neutrons deposit energy in the ZDC) and to peripheral collisions (where most spectator neutrons are bound in larger nuclear fragments and are thus deflected away from the ZDC by the RHIC magnets), respectively. 

\begin{figure}
    \centering
        \includegraphics[height=3.0in]{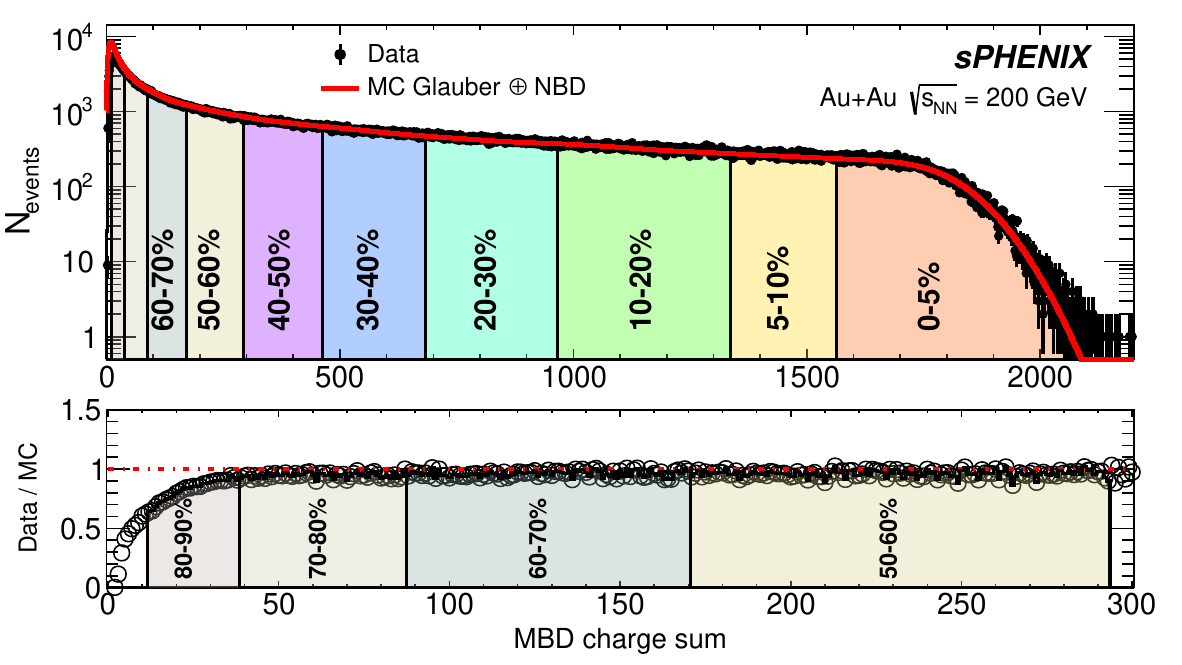}
        \caption{The MBD charge sum distribution in minimum-bias \auau{} events (points), compared to the best fit from a MC Glauber $\oplus$ NBD simulation (red line). The shaded bands indicate  5\%-wide centrality intervals. The lower panel highlights the ratio of the MBD charge sum to the MC Glauber $\oplus$ NBD fit in the peripheral event region, showing the effect of the trigger turn on.  The MBD charge sum is in units of calibrated MIPs.}
        \label{fig:mbdcent}
\end{figure}

The MBD charge sum is used in each event to assign a centrality value that characterizes the level of geometric overlap between the two colliding nuclei. Centrality percentiles are derived by fitting the MBD charge sum distribution for all MB events to a model of the particle production and event sampling based on the convolution of a MC Glauber simulation~\cite{Loizides_2015} with a negative binomial distribution (MC Glauber $\oplus$ NBD)~\cite{PHENIX:2004vdg,phenix_centrality_2}.  Figure~\ref{fig:mbdcent} shows the MBD charge sum distribution, along with the best fit of the MC Glauber $\oplus$ NBD model and resulting centrality intervals. The bottom panel of Figure~\ref{fig:mbdcent} shows the peripheral event region to highlight MB trigger efficiency turn-on region. The total MB selection efficiency for inelastic \auau events is then determined by comparing the integral of the data distribution to the integral of the MC Glauber $\oplus$ NBD, normalized for MBD charge sum $>$ 150. The resulting efficiency is $93.6\%^{+3.4\%}_{-3.1\%}$, similar to previous \auau data-taking in PHENIX where the MBD was in a different location~\cite{PHENIX:2004vdg}.
The Glauber parameters characterizing the centrality intervals and the extracted \Npart values for these centrality intervals are shown in Table~\ref{tab:glauber}.

Since the calorimeter towers are projective in $\eta$ from the nominal center of the detector at $z = 0$, events used in the offline analysis were required to have a $z$-vertex position, reconstructed by the MBD detector, within $10$~cm of this point.
Events were further selected to be in the centrality range of $0$--$70$\% to ensure complete minimum-bias event selection efficiency, while also including a broad range of geometric sizes of the produced QGP. After the above selection requirements, the analyzed dataset comprised 518,000 events within the $0$--$70$\% centrality range. 

\begin{table}
\begin{subfigure}[t]{0.45\textwidth}
\centering
    \renewcommand{\arraystretch}{1.5}
    \begin{tabular}{cc}
    Glauber Parameter & Value \\
    \hline
    $\sigma_{NN}$ [mb] & $42 \pm 3$ \\
    Nuclear radius [fm] & $6.38_{-0.13}^{+0.27}$ \\
    Skin depth [fm]  & $0.535_{-0.010}^{+0.020}$ 
    \end{tabular}
\end{subfigure}
\begin{subfigure}[t]{0.45\textwidth}
\centering
    \renewcommand{\arraystretch}{1.3}
 \begin{tabular}{cc}
    Centrality interval & \Npart \\
    \hline
        0-5\% & $350.0 \pm 2.2$ \\
        5-10\% & $301.8 \pm 3.2$ \\
        10-20\% & $238.1 \pm 4.1$ \\
        20-30\% & $170.8 \pm 4.9$ \\
        30-40\% & $118.5 \pm 5.2$ \\
        40-50\% & $78.3 \pm 5.1$ \\
        50-60\% & $47.8 \pm 4.5$ \\
        60-70\% & $26.1 \pm 3.6$ \\
    \end{tabular}
\end{subfigure}
\caption{Glauber model parameters (left) for \auau{} collisions at \sqsntwo. Centrality intervals and average number of participating nucleons ($\Npart$) for \auau collisions at \sqsntwo (right) obtained using Glauber model.}
\label{tab:glauber}
\end{table}


\subsection{Calorimeter calibration}

\begin{figure}
    \centering
        \begin{subfigure}[t]{0.5\textwidth}
        \centering
        \includegraphics[width=\textwidth]{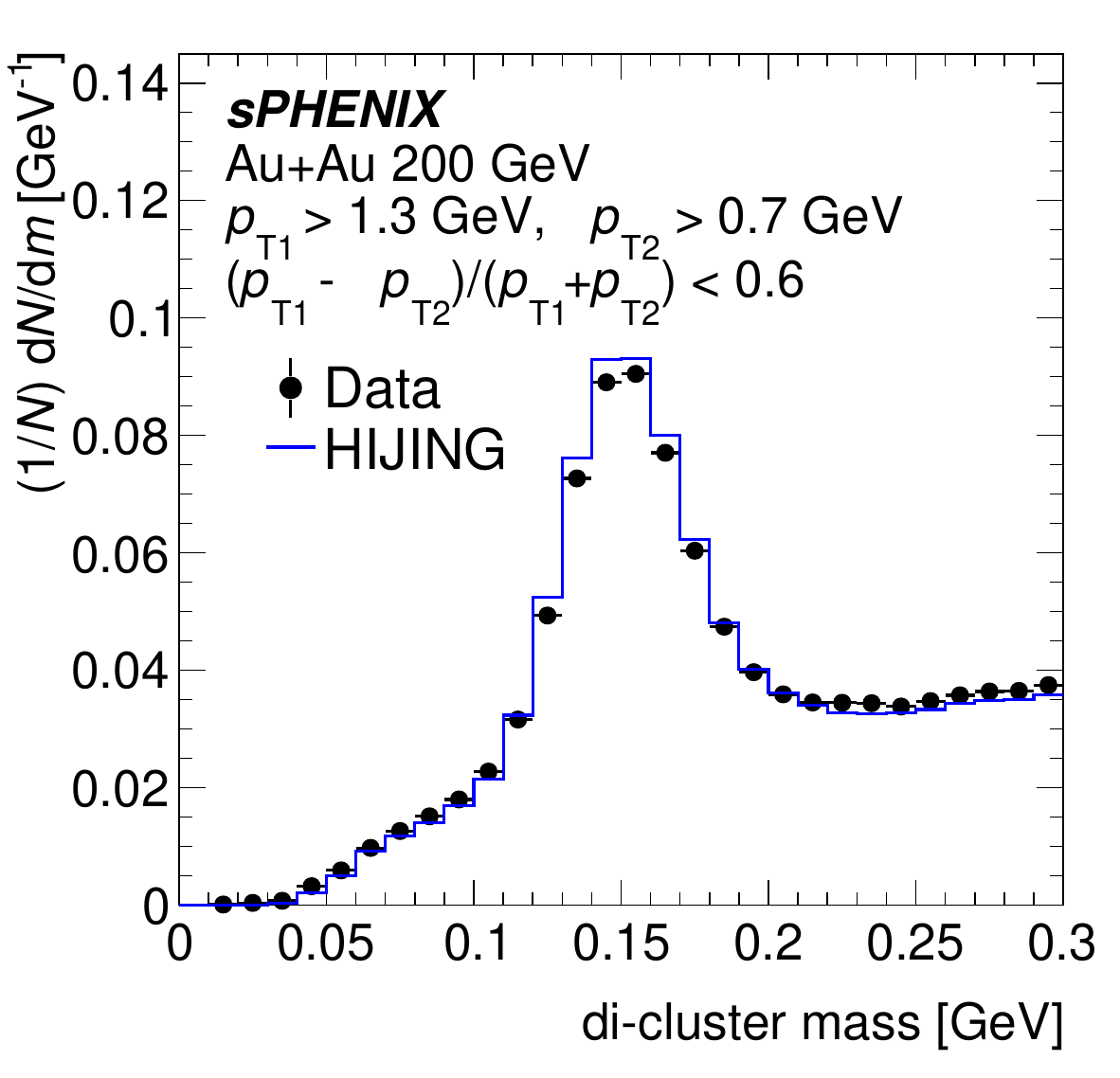}
        \caption{}
        \label{fig:emcal_pi0}
    \end{subfigure}%
    ~ 
    \begin{subfigure}[t]{0.5\textwidth}
        \centering
        \includegraphics[width=\textwidth]{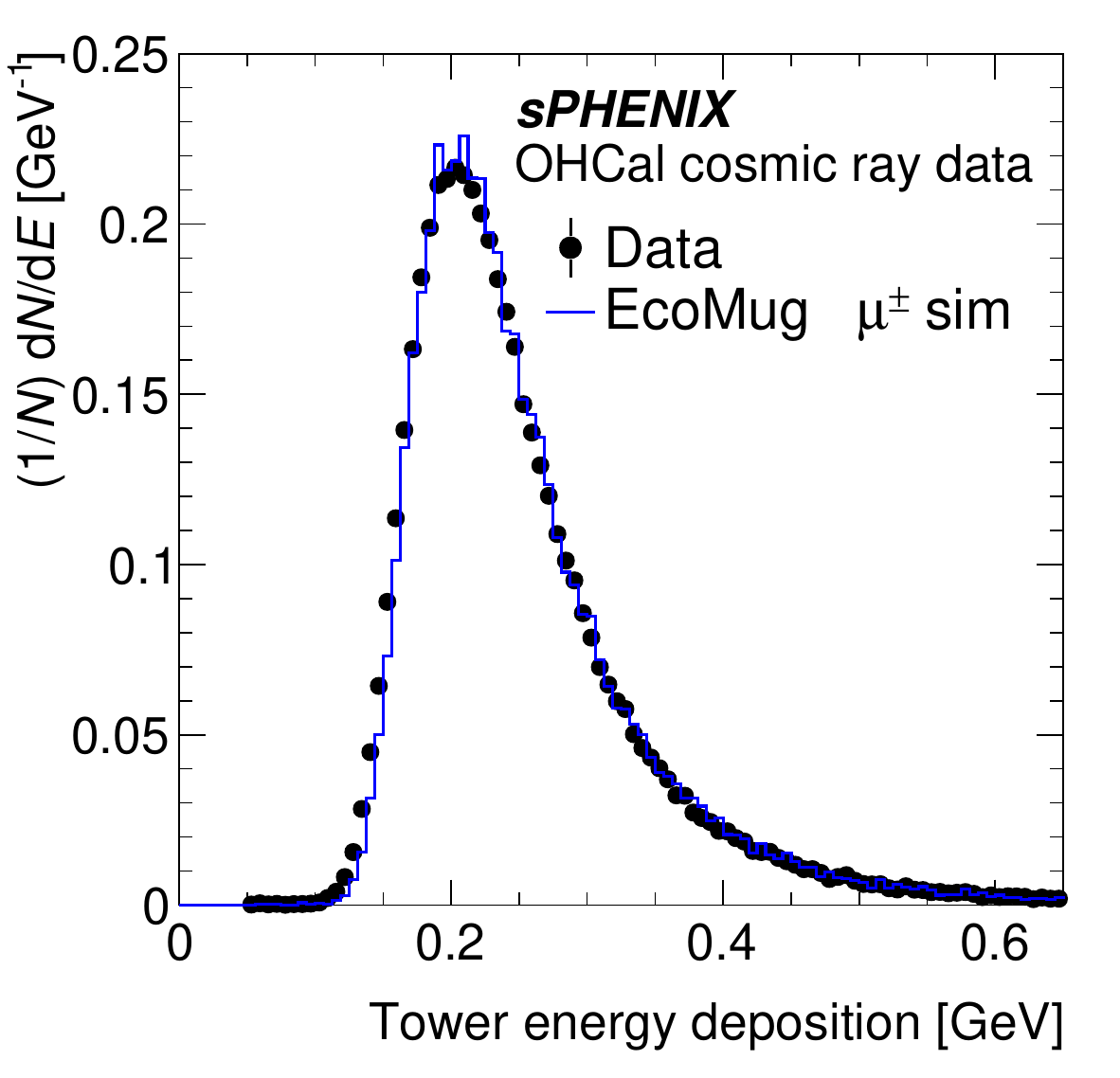}
        \caption{}
        \label{fig:hcal_mip}
    \end{subfigure}
    \cprotect\caption{Panel (a) provides an example of a reconstructed EMCal di-cluster invariant mass distribution, similar to those used for {\it{in situ}} EMCal tower calibrations. The distributions are made from EMCal cluster pairs using Run 2024 \auau{} data (points) and a \textsc{geant-4} simulation of  \verb|HIJING| events (histogram). The prominent peak arises from $\pi^{0}\to\gamma\gamma$ decays. Panel (b) illustrates an example of the measured energy distribution in a single OHCal tower, comparing the MIP distribution from cosmic-ray data from the detector (points) and from a \textsc{geant-4} simulation of cosmic-ray muons from EcoMug (histogram).}
\end{figure}


For the reconstruction of calorimeter energy deposits, time samples from the calorimeter electronics are fit to a waveform template derived from Run 2024 beam data to extract the analog-to-digital converter (ADC) signal amplitude in each tower in each event. In addition to the hardware-level zero-suppression threshold set at 2$\sigma$ of the calorimeter pedestal noise from the start of Run 2024, residual low-energy noise close to but above this noise limit are zero-suppressed offline using a peak-minus-pedestal algorithm, with different ADC thresholds for each of the three calorimeter subsystems to account for small increases in calorimeter noise over the full length of Run 2024. These thresholds are chosen using pedestal-only runs taken during periods of no beam directly prior the analysis beam runs to ensure that the contributions to the \detdeta measurement from noise are negligible.

For the EMCal, the absolute energy scale calibration is established using an $\eta$-dependent calibration of the $\pi^{0}\to\gamma\gamma$ mass peak in data to the same position as in simulation using the runs within this measurement's dataset. A relative calibration is first applied to EMCal towers enforcing the expected $\phi$ symmetry to eliminate tower-by-tower gain differences, followed by an absolute $\eta$-dependent calibration for each EMCal $\eta$ slice to set the position of the $\pi^{0}$ mass peak. EMCal clusters used for this absolute calibration are formed by grouping contiguous EMCal towers with energies exceeding 70~\MeV. An 11.9\% smearing on the EMCal cluster energies in simulation is necessary to match the relative width of the $\pi^{0}\to\gamma\gamma$ mass peak in data. Figure~\ref{fig:emcal_pi0} compares the distribution of EMCal di-cluster masses in \auau{} data and simulated events with an example kinematic selection, demonstrating the agreement in the $\pi^{0}$ mass peak position and width.   The $\pi^{0}$ mass peak position is shifted in both data and simulation from the Particle Data Group value~\cite{ParticleDataGroup:2024cfk} due to a convolution of the falling energy spectrum and the resolution in the energy response.


The IHCal and OHCal absolute energy scale calibrations are performed using the MIP energy depositions from cosmic ray muons from data taken during no-beam periods of Run 2024. Since these calibration runs occur throughout the data-taking period, the effect of small, time-dependent gain variations, which arise from local detector conditions, such as changes in temperature, is negligible for the calibrations used for this measurement. A single HCal tower trigger is utilized to collect a sample of cosmic ray muon events using an energy threshold well below the HCal characteristic MIP peak. The detector response to cosmic ray muons is simulated using EcoMug~\cite{ecomug} to generate a cosmic ray muon flux incident on the sPHENIX detector and propagated through a \textsc{geant-4}~\cite{GEANT4} simulation to accurately model the angular distribution of cosmic rays seen by the sPHENIX HCals. A set of offline cuts consisting of minimum and maximum requirements on the energy in surrounding HCal towers, applied to both simulation and data, is used to select a set of muon events which propagate through the HCal tower in the azimuthal direction and deposit energy in a majority of the tower's scintillating tiles. This configuration results in a narrow, characteristic MIP energy peak needed for a precise calibration. Figure~\ref{fig:hcal_mip} shows a comparison of the MIP energy distribution for an example OHCal tower from cosmic ray data and the simulation described above. An additional factor is then applied in both data and simulation to correct the measured scintillator MIP energy distribution to an estimate of the energy deposited in both the scintillator and absorber material. This factor is determined using the average response in the full HCal detector (scintillating tiles and absorber material) for high-$E_\mathrm{T}$ single hadrons determined in simulation as validated by test beam studies~\cite{EMCal_HCal_test_beam}.

After calibration, the detector-level \detdeta in each calorimeter system was reconstructed for each centrality class via:
$E_{T}(\eta) = \sum_{i=1}^{N_{\textrm{towers}} } E_{\mathrm{tower}}(\eta) \text{sin}(\theta_{\mathrm{tower}}(\eta))$
where the polar angle, $\theta$, and $\eta$ of each tower is determined with respect to the vertex position.

\subsection{Detector response correction factors}

Simulations of \auau events are produced using three MC event generators, \verb|HIJING|~\cite{hijing}, \verb|AMPT|~\cite{ampt}, and \verb|EPOS4|~\cite{epos}. These simulations are used in the analysis to derive correction factors for the reconstructed \detdeta values measured in the sPHENIX calorimeter measurements.  
These corrections include accounting for charged particles with transverse momentum \pt $\lesssim 180$~\MeV{} that curl up in the sPHENIX magnetic field before reaching the calorimeters.

\verb|HIJING| is a high-energy, heavy-ion and proton-proton collision event generator that uses an MC Glauber model of the nucleus-nucleus collision geometry and perturbative QCD to model hard scatterings as parton mini-jets. The model includes multi-parton interactions, and initial and final state radiation effects. \verb|HIJING| uses the Dual Parton Model for soft interactions and the Lund string model for hadronization. The \verb|AMPT| generator uses \verb|HIJING| to generate the initial parton distributions in each event, which are then evolved through a parton cascade of $2\to2$ elastic collisions, followed by hadronization using either the Lund string model or a quark coalescence model. The resulting hadron distribution is then developed through a hadron transport model including elastic and inelastic scatterings, and resonance decays. The \verb|EPOS4| generator models collisions using an $S$-matrix approach to Gribov-Regge Theory~\cite{Drescher_2001}, which simultaneously describes soft and hard scatterings. After primary interactions are developed based on this $S$-matrix approach for parallel scatterings, the system undergoes a core-corona separation procedure, followed by hydrodynamic evolution, hadronization, and hadronic transport. 

The simulations described above are used to correct the reconstructed energy for the response of the calorimeters. The generator-level particle spectra are reweighted to match previous measurements of identified particles in \auau collisions at \sqsntwo at RHIC. The weights for $\pi^{\pm}$, $K^{\pm}$, protons, and neutrons are derived from PHENIX data~\cite{phenix_particle_spectra}, with the $\pi^{0}$ and $K^{0}$ weights set to the average of their charged counterparts. The weights for $\Lambda^{0}$ and $\Sigma^{\pm}$ baryons are derived from STAR data~\cite{star_lambda_spectra}, and the weights of all other baryons were set to those of protons and neutrons. These weights are determined as a function of $p_{T}$, separately for each centrality interval. Outside the $p_T$ range of the previous measurements, a constant weight equal to the last value within that range is used. An additional crosscheck using identified particle spectra differential in both $p_{T}$ and rapidity using BRAHMS data~\cite{Brahms_pi_spectra,Brahms_p_spectra} is also performed and the variation in the MC correction factors from using rapidity dependent reweighting factors is included as a systematic uncertainty on the MC correction factors.
This reweighing procedure is applied to the correction factors from \verb|EPOS4| to determine the nominal values, with the comparisons to the other models used to evaluate the uncertainty in the predictions from the physics modeling.



Events from each of these MC generators are simulated within a \textsc{geant-4} description of the detector geometry followed by the same reconstruction, calibration, and analysis chain as the data.
The simulations are then used to determine multiplicative correction factors to correct the measured detector-level \detdeta to a particle-level, or ``true'', \detdeta. 
First, the particle-level $\Sigma E_{T}$ is calculated by summing the $E_{T,\mathrm{particle}}$ for all final-state particles within the detector's nominal acceptance as a function of $\eta$. In the sum, baryons contribute only their kinetic energy, antibaryons contribute their kinetic energy plus 2$m_{N}$ (i.e., their total energy plus $m_N$), where $m_{N}$ is the nucleon mass, and all other particles contribute their total energy. 
The reconstructed-level $\Sigma E_{T}$ in simulation is determined from calorimeter tower energies and positions with the same analysis chain as used on real data. 
Correction factors are determined for each calorimeter sub-system and each centrality bin by using the mean reconstructed $\Sigma E_{T}$ in the specific calorimeter divided by the particle-level value for all particles: $C(\eta) = \sum E_{T,\mathrm{tower}}(\eta)/\sum E_{T,\mathrm{particle}}(\eta)$.

In this analysis, the EMCal and HCal (combined IHCal + OHCal) are each used to make separate measurements of \detdeta, and the EMCal and HCal are used together for a full calorimeter measurement. The correction factors are approximately 0.7 for the EMCal, 0.2 for the HCal, and 0.9 for the full calorimeter. The relative magnitude of these correction factors reflect the fraction of the produced energy seen by the EMCal versus HCal, where the EMCal in front of the HCal receives the EM and a large fraction of the hadronic energy of the collision. These correction factors account for the effect of the magnetic field, small areas of the calorimeters that are inactive during data-taking, particle in-flow and out-flow from the detector acceptance, losses in the magnet region between the IHCal and OHCal, and do not depend strongly on centrality. 

\section{Systematic uncertainties}

Major sources of systematic uncertainty in the measurement include the calorimeter energy calibration, the calorimeter response to hadrons, the modeling of particle spectra in simulation, choice of zero suppression thresholds in the calorimeter signal extraction, detector acceptance, and $z$-vertex resolution effects. These sources are summarized below. Table~\ref{table:SystUncertainties} shows the $\eta$-averaged variation of the magnitude contributed from each of the uncertainty sources to the measurement as well as the total uncertainty for different event centralities. The systematic uncertainties for all measurements are strongly correlated in $\eta$ and the statistical uncertainties are negligible compared to the systematic uncertainties.




\begin{table}[t]
\centering
\begin{tabular}{ lccc }
Uncertainty Source [\%] & EMCal-Only & HCal-Only & Full Calorimeter \\
\hline
Calibration & 2.6 & 2.7 & 2.1 \\
Hadronic response & 4.1 & 6.6 & 4.7 \\
Modeling & 1.4--2.1 & 2.5--3.8 & 1.6--2.2 \\
Zero suppression thres. & 1.0--5.8 & 0.2--0.3 & 0.8--4.4 \\
$z$-vertex resolution & 0.3--0.4 & 0.1--0.2 & 0.2--0.3 \\
Acceptance & 0.2--0.5 & 0.2--0.6 & 0.1--0.3 \\
Total & 5.3--8.1 & 7.7--8.3 & 5.6--7.4 \\

\end{tabular}
\caption{Overview of major systematic uncertainties contributing to the measurement. The range of magnitudes of uncertainties, in percent, are shown for each source (rows) for the measurements using different calorimeter systems (columns), with the total uncertainty shown in the final row. The ranges correspond to the $\eta$-averaged variation of the magnitude for different event centralities. Uncertainties on \Npart depend only on centrality, range from $0.6$--$13.8$\%, and are listed in Table~\ref{tab:glauber}.}
\label{table:SystUncertainties}
\end{table}

The calibration uncertainties arise from the matching of the $\pi^0\to\gamma\gamma$ mass distribution (cosmic muon MIP distribution) measured in the EMCal (HCals) between simulation and data, including residual data-simulation differences, potential variations in the calibration in different regions of the detector, and statistical uncertainties on the tower-by-tower calibrations in data.
The hadronic response uncertainty is evaluated by comparing the agreement of the single hadron response between data and simulation in beam tests of prototypes of the sPHENIX calorimeter system~\cite{EMCal_HCal_test_beam}. This is the dominant uncertainty, reaching nearly 7\% for the HCal-only measurement. To determine the hadronic response uncertainty for the EMCal-only and full calorimeter measurements, which are comprised of a combination of EM and hadronic energy, the hadronic response uncertainty was only applied to the fraction of hadronic energy included in each of the measurements determined from the simulations used to determine the analysis correction factors. 

The sensitivity to the physics modeling is evaluated by deriving the correction factors using the reweighted $\verb|HIJING|$ or $\verb|AMPT|$ simulations rather than $\verb|EPOS4|$, and by considering alternative $\eta$-dependent particle spectra as suggested by BRAHMS data~\cite{Brahms_p_spectra,Brahms_pi_spectra}. This is the second-largest uncertainty source for the HCal-only measurement. The uncertainty from the reconstruction of calorimeter towers consistent with noise is evaluated by varying the zero-suppression threshold applied offline and instead performing the full waveform-template fit for these low-energy towers. This uncertainty is most significant for the EMCal in peripheral events, where the fraction of calorimeter towers with real energy deposits is much lower than that in central events.
The impact from the finite $z$-vertex position resolution is conservatively estimated by artificially shifting the reconstructed $z$-vertex position by $\pm3$~cm. Finally, an uncertainty related to the stability of the detector acceptance and local conditions over time is evaluated by repeating the analysis for different recorded \auau runs. These last two effects ($z$-vertex resolution and acceptance) are sub-dominant compared to the other sources.

Uncertainties on the extracted \Npart{} values are evaluated using a standard set of variations in the MC Glauber modeling and centrality determination. These include varying the nucleon--nucleon cross-section and other geometric parameters in the Glauber model and varying centile cuts according to the uncertainty in the total efficiency listed in Table~\ref{tab:glauber}. The dominant source of uncertainty for the \Npart{} values is the MB trigger inefficiency. For the centrality intervals used in this measurement, the uncertainties range from $0.6$\% in $0$--$5$\% events to $13.8$\% in $60$--$70$\% events. The \Npart{} uncertainties only contribute to the measurement of $\detdeta / (0.5$\Npart).

\section{Results}
\label{sec:results}
\begin{figure}[t]
    \centering
    \includegraphics[width=\linewidth]{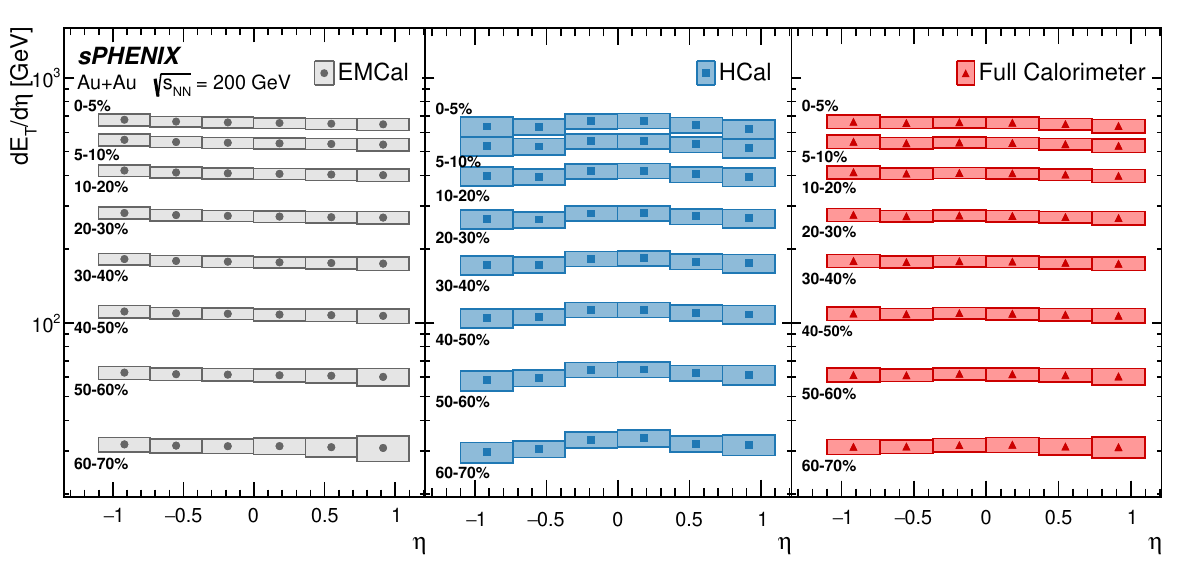}
    \caption{Summary of \detdeta measurements in Au+Au collisions at $\sqrt{s_\mathrm{_{NN}}} = 200$~GeV over the pseudorapidity range $\left|\eta\right| < 1.1$. Different data series correspond to centrality selections within the range $0$--$70$\% (labeled). The different panels show the results determined using the EMCal-only (left), HCal-only (center), and the full calorimeter system (right). The vertical size of the boxes around each point indicate the total systematic uncertainty, while the horizontal size indicates the $\eta$ bin width.}
    \label{fig:subsystem_detdeta}
\end{figure}

Results for \detdeta are presented in Figure~\ref{fig:subsystem_detdeta} as a function of $\eta$ and for various centrality intervals within the range $0$--$70$\%. 
The \detdeta values have a strong dependence on centrality, varying by more than a factor of 10 between the most peripheral and most central events considered here. 
The measurements using each of the three methods (EMCal-only, HCal-only, and the full calorimeter system) are shown, and the results are compatible between them. 
No significant dependence on $\eta$ is observed within the range $\left|\eta\right| < 1.1$.
The measurements at the same absolute value of $\eta$ are generally consistent across all methods and centralities. Small deviations, up to a few percent, are observed at forward-$\left|\eta\right|$ in the most central intervals and are accounted for within the total estimated uncertainties.

\begin{figure}[h]
    \centering
    \includegraphics[height=3.6in]{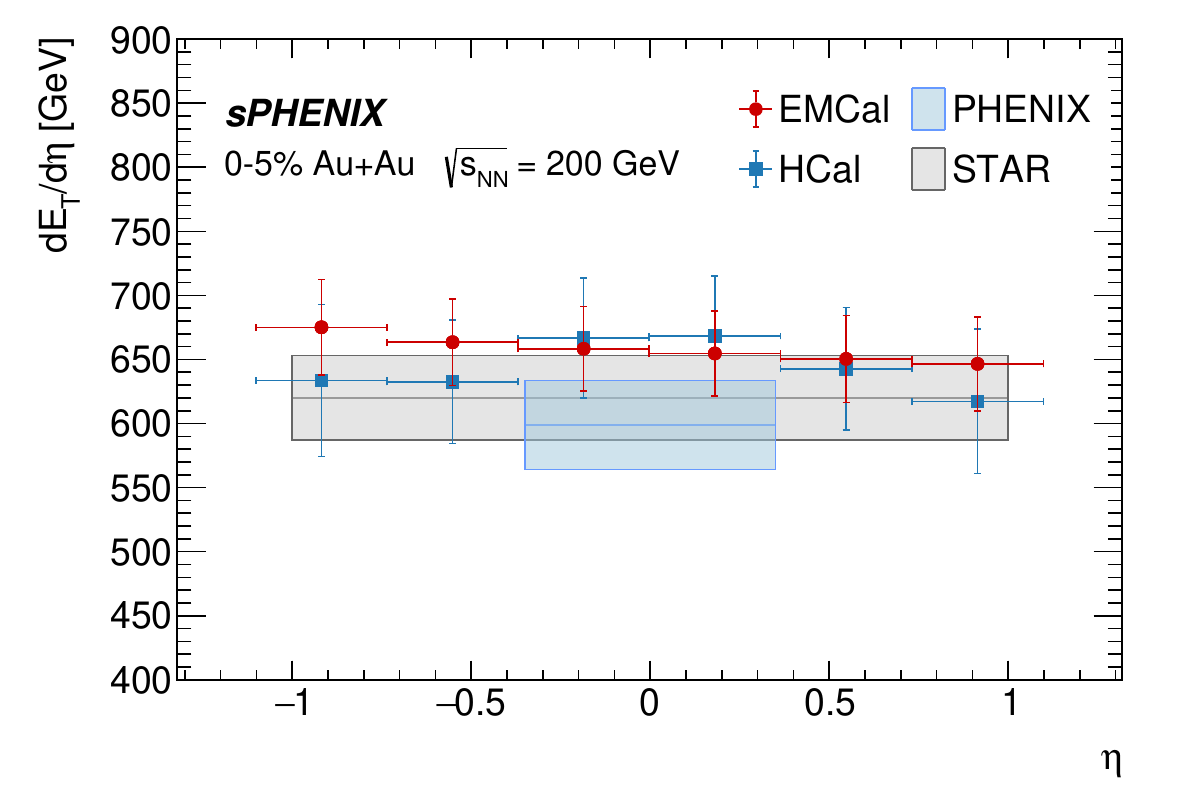}
    \cprotect\caption{Measurement of \detdeta in 0--5\% central \auau{} events using only EMCal (red points) and only HCal (blue points), as a function of $\eta$ over the range $\left|\eta\right| < 1.1$. The vertical error bars show the total systematic uncertainty. The measurements from PHENIX~\cite{PHENIX:2004vdg} (blue box) and STAR~\cite{STAR_detdeta} (gray box) in this centrality interval are shown for comparison, with the vertical and horizontal size of the box indicating the total uncertainty and the measurement range in $\eta$, respectively.}
    \label{fig:emcal_hcal_comp}
\end{figure}

\begin{figure}[h]
    \centering
    \includegraphics[height=3.6in]{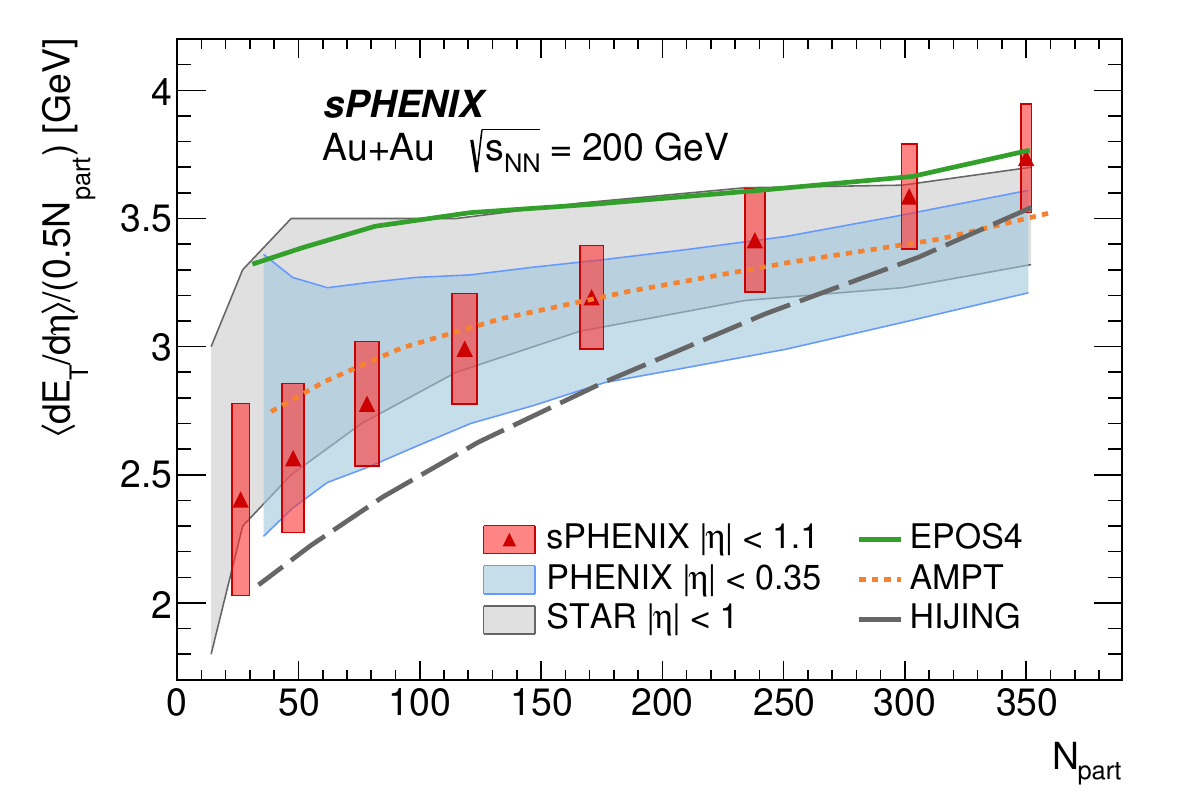}
    \cprotect\caption{Measured \detdeta normalized by the estimated number of participant pairs ($0.5$\Npart), as a function of \Npart, using the full calorimeter system (red triangles). The vertical size of the boxes indicates the total uncertainty, which includes the uncertainty in \Npart. Measurements by PHENIX~\cite{PHENIX:2004vdg} (blue band) and STAR~\cite{STAR_detdeta} (gray band) are shown for comparison, with the vertical size of the band indicating the uncertainty range. MC event generator predictions are shown for \verb|EPOS4| (solid green line), \verb|AMPT| (dotted orange line), and \verb|HIJING| (dashed black line).}
    \label{fig:detdeta_vs_npart}
\end{figure}

In Figure~\ref{fig:emcal_hcal_comp}, the EMCal-only and HCal-only \detdeta measurements in the most central 0--5\% Au+Au events are overlaid to highlight their agreement as the EMCal and HCal are sensitive to the deposited energy from different particles and have independent calibration sequences. 
Fig~\ref{fig:emcal_hcal_comp} includes a comparison to previous PHENIX~\cite{PHENIX:2004vdg} and STAR~\cite{STAR_detdeta} results in this centrality interval. For comparison, the STAR results measured from $0 < \eta < 1$ are symmetrized over range $|\eta| < 1$. Both the sPHENIX EMCal-only and HCal-only measurements are compatible with the previous measurements at RHIC performed using only electromagnetic calorimetry or a combination of tracking detectors and an electromagnetic calorimeter, while providing additional granularity in $\eta$.
For this centrality range, the total uncertainties are comparable between the sPHENIX, PHENIX, and STAR measurements. The uncertainties for the sPHENIX EMCal-only results are similar in magnitude to that for the previous PHENIX results, which also used only an EM calorimeter for the \detdeta measurement.

Figure~\ref{fig:detdeta_vs_npart} shows the average $\detdeta$, normalized per participant pair as a function \Npart, measured using the full sPHENIX calorimeter system. The  $\langle \detdeta\rangle/(0.5 \Npart)$ values gradually increase from approximately 2.5 to 3.5 GeV per participant pair over the reported \Npart range. The results are compared to previous measurements by PHENIX~\cite{PHENIX:2004vdg} and STAR~\cite{STAR_detdeta}. The sPHENIX measurement is compatible with these and features an improved precision in peripheral events.
Figure~\ref{fig:detdeta_vs_npart} also includes comparisons to the \detdeta per participant pair in generator-level \verb|HIJING|, \verb|EPOS4|, and \verb|AMPT| events (each without the applied reweighting used for analysis correction factors). Across the full \Npart range, \verb|AMPT| best describes both the overall magnitude and the dependence on \Npart in the sPHENIX measurement.
\section{Summary}
\label{sec:summary}
This paper presents a measurement of the transverse energy per unit pseudorapidity (\detdeta) in \auau collisions at \sqsntwo performed with the sPHENIX calorimeter system over the range $\left|\eta\right| < 1.1$. This constitutes the first measurement of this observable performed with a hadronic calorimeter at RHIC. The measurement is first performed using only the EMCal and only the HCal, with good agreement between the two despite the different sensitivities and calibration procedures for these detector systems. The measurement is performed with the full calorimeter system, differentially in centrality over the range $0$--$70$\%, and is also reported in ratio to the estimated number of participant pairs. The sPHENIX measurement is compatible with previous measurements from PHENIX and STAR, with the present result providing additional granularity in $\eta$ and improved uncertainties in peripheral events. The measurement is also used to benchmark the predictions for this observable in a number of MC event generators. This measurement demonstrates the performance of the different elements of the sPHENIX calorimeter system over a large dynamic range, and is a necessary step towards the envisioned physics program of calorimeter jet measurements performed with the detector.

\section*{Acknowledgements}

We thank the Collider-Accelerator Division, SDCC and Physics
Departments at Brookhaven National Laboratory and the staff of the
other sPHENIX participating institutions for their vital
contributions. We acknowledge support from the Office of Nuclear
Physics and Graduate Student Research (SCGSR) program in the Office of
Science of the U.S. Department of Energy, the U.S. National Science
Foundation, the National Science and Technology Council, the Ministry
of Education of Taiwan, and the Ministry of Economic Affairs (Taiwan),
the Ministry of Education, Culture, Sports, Science, and Technology
and the Japan Society for the Promotion of Science (Japan), Basic
Science Research Programs through NRF funded by the Ministry of
Education and the Ministry of Science and ICT (Korea) and the Swedish
Research Council, VR (Sweden).

\bibliographystyle{unsrturl}
\bibliography{bib/references}

\newpage

\appendix

\section*{The sPHENIX Collaboration}

\begin{flushleft}
\small

M.~I.~Abdulhamid$^{13}$,
U.~Acharya\,\orcidlink{0000-0001-8560-963X}\,$^{13}$,
E. R.~Adams$^{7}$,
G.~Adawi$^{13}$,
C.~A.~Aidala\,\orcidlink{0000-0001-9540-4988}\,$^{26}$,
Y.~Akiba$^{39}$,
M.~Alfred$^{14}$,
S.~Ali$^{13}$,
A.~Alsayegh$^{10}$,
S.~Altaf$^{15}$
% ...

\section*{Collaboration Institutes}

$^{1}$ Augustana University, Sioux Falls, South Dakota\\
$^{2}$ Baruch College, City University of New York, New York, New York\\
$^{3}$ Brookhaven National Laboratory, Upton, New York\\
$^{4}$ University of California, Los Angeles, California\\
$^{5}$ Central China Normal University, Wuhan, Hubei\\
$^{6}$ IRFU, CEA, Universit\'{e} Paris-Saclay, Gif-sur-Yvette, France\\
$^{7}$ University of Colorado, Boulder, Colorado\\
$^{8}$ Columbia University, New York, New York\\
$^{9}$ Debrecen University, Debrecen, Hungary\\
$^{10}$ Florida Agricultural and Mechanical University, Tallahassee, Florida\\
$^{11}$ Florida State University, Tallahassee, Florida\\
$^{12}$ Fudan University, Shanghai\\
$^{13}$ Georgia State University, Atlanta, Georgia\\
$^{14}$ Howard University, Washington, District of Columbia\\
$^{15}$ University of Illinois at Urbana-Champaign, Urbana, Illinois\\
$^{16}$ Institute for Information Industry, Taipei\\
$^{17}$ Iowa State University, Ames, Iowa\\
$^{18}$ Japan Atomic Energy Agency, Naka, Ibaraki, Japan\\
$^{19}$ Korea University, Seoul, Korea\\
$^{20}$ Lawrence Berkeley National Laboratory, Berkeley, California\\
$^{21}$ Lawrence Livermore National Laboratory, Livermore, California\\
$^{22}$ Lehigh University, Bethlehem, Pennsylvania\\
$^{23}$ Los Alamos National Laboratory, Los Alamos, New Mexico\\
$^{24}$ Lund University, Lund, Sweden\\
$^{25}$ Massachusetts Institute of Technology, Cambridge, Massachusetts\\
$^{26}$ University of Michigan, Ann Arbor, Michigan\\
$^{27}$ Morgan State University, Baltimore, Maryland\\
$^{28}$ Muhlenberg College, Allentown, Pennsylvania\\
$^{29}$ Nara Women's University, Nara, Nara, Japan\\
$^{30}$ National Central University, Taoyuan City\\
$^{31}$ National Research Nuclear University, MEPhI, Moscow Engineering Physics Institute, Moscow, Russia\\
$^{32}$ National Taiwan University, Taipei\\
$^{33}$ University of North Carolina, Greensboro, North Carolina\\
$^{34}$ Oak Ridge National Laboratory, Oak Ridge, Tennessee\\
$^{35}$ Ohio University, Athens, Ohio\\
$^{36}$ Peking University, Beijing\\
$^{37}$ Purdue University, West Lafayette, Indiana\\
$^{38}$ Pusan National University, Pusan, Korea\\
$^{39}$ RIKEN Nishina Center for Accelerator-Based Science, Wako, Saitama, Japan\\
$^{40}$ RIKEN BNL Research Center, Brookhaven National Laboratory, Upton, New York\\
$^{41}$ Rikkyo University, Toshima, Tokyo, Japan\\
$^{42}$ Rutgers University, Piscataway, New Jersey\\
$^{43}$ State University of New York, Stony Brook, New York\\
$^{44}$ Sun Yat-sen University, Guangzhou, Guangdong\\
$^{45}$ University of Tennessee, Knoxville, Tennessee\\
$^{46}$ University of Texas, Austin, Texas\\
$^{47}$ Vanderbilt University, Nashville, Tennessee\\
$^{48}$ Wayne State University, Detroit, Michigan\\
$^{49}$ Weizmann Institute of Science, Rehovot, Israel\\
$^{50}$ Yonsei University, Seoul, Korea\\
$^{51}$ United States Merchant Marine Academy, Kings Point, New York\\

\end{flushleft}

\end{document}